\documentclass[%
reprint,
 amssymb, amsmath,%
 aps,
 prl,
]{revtex4-1}

\usepackage[pdftex]{graphicx}
\graphicspath{{./../figures/}}
\DeclareGraphicsExtensions{.pdf,.jpeg,.png}

\usepackage{stmaryrd}
\usepackage{docs}%
\usepackage{bm}%
\usepackage[colorlinks=true,linkcolor=blue]{hyperref}%
\expandafter\ifx\csname package@font\endcsname\relax\else
 \expandafter\expandafter
 \expandafter\usepackage
 \expandafter\expandafter
 \expandafter{\csname package@font\endcsname}%
\fi
\begin{document}

\title{%
       Characterization of the individual short-term frequency stability of Cryogenic Sapphire Oscillators at the $10^{-16}$ level.}

\author{Christophe Fluhr$^{1}$, Serge Grop$^{2}$, Beno\^{i}t Dubois$^{3}$, Yann Kersal\'e$^{1}$,  Enrico Rubiola$^{1}$ and Vincent Giordano$^{1}$}
\email{giordano@femto-st.fr}

\affiliation{\vskip 5mm $^{1}$ FEMTO-ST Institute; UMR 6174 (CNRS), 26 Chemin de l'\'Epitaphe, 25000 Besan\c{c}on -- FRANCE \\
 $^{2}$ Alemnis GmbH, Feuerwerkerstrasse 39, 3602 Thun -- SWITZERLAND\\
$^{3}$ FEMTO Engineering, 32 avenue de l'Observatoire, 25000 Besan\c{c}on -- FRANCE\\ }

\vskip 5mm
\date{2015 December, 10}

\begin{abstract}
We present the characterisation of three Cryogenic Sapphire Oscillators using the three-corner-hat method. Easily implemented with commercial components and instruments, this method reveals itself very useful to analyse the frequency stability limitations of these state-of-the-art ultra-stable oscillators. The best unit presents a fractional frequency stability better than $5\times 10^{-16}$ at 1 s and below $2\times 10^{-16}$ for $\tau<5,000$ s.
\end{abstract}

\maketitle


\section{Introduction}

The Cryogenic Sapphire Oscillator (CSO) based on a sapphire whispering gallery mode resonator cooled near 6 K is currently the microwave signal source presenting the highest short-term frequency stability
for integration times  $\tau=1\ldots10,000 $ s.  With a fractional frequency stability better than $1\times 10^{-15}$, the CSO allows the operation of the laser-cooled microwave atomic clocks at the quantum limit \cite{santarelli99-prl}. It provides the means to improve the resolution of the space vehicles ranging and Doppler tracking provided by Deep Space Networks as well as those of Very Long Baseline Interferometry (VLBI) Observatories  \cite{dick95,rsi10-elisa,ell11-elisa,nand2011-mtt,doeleman2011,rioja2012}. The CSO can also enhance the calibration capability of Metrological Institutes or help the qualification of high performances clocks or oscillators \cite{santarelli99-prl, AppPhysB-2014, takamizawa2014}. The Oscillator-IMP project funded in the frame of the french national \it{Projets d'Investivement d'Avenir}\rm\ (PIA) targets at being a facility dedicated to the measurement of noise and short-term stability of oscillators and devices in the whole radio spectrum (from MHz to THz), including microwave photonics. The Oscillator-IMP platform consists of highly sophisticated instruments for the measurement of short-term stability and spectral purity, and of the most stable frequency references: atomic clocks and ultra-stable oscillators. Among them, we are implementing a set of three CSOs. \\

Fractional  frequency stabilities better than $1\times 10^{-15}$ have already been demonstrated by beating two nearly identical CSOs \cite{RSI-2012, hartnett-2012-apl}. The Allan standard Deviation (ADEV) of one unit is simply obtained by substracting 3 dB to the actual result, assuming the two oscillator noises are equivalent and uncorrelated. This assumption is not warranted especially when developping a new instrument at the state-of-the-art. The recent implementation of three CSOs in our lab was the opportunity to test the three-cornered-hat method \cite{gray74} to extract the individual ADEV. Very preliminary measurements based on these method have been presented in \cite{ifcs-2015-cso}. Since one of the CSO has been improved and more data have been accumulated. The current results demonstrate the capability of the method, which has been simply implemented with commercial components and counters. Althought based on the same general configuration, the three CSOs lightly differ from each other. Different thermal configurations have been tested and only one resonator is completly optimized, the two others recently implemented still need some adjustments. The three-cornered-hat method gives us information about each CSO and thus will help us to optimize its functioning.\\

\section{CSO description}

\begin{figure}[ht!]
	\centering
	\includegraphics[width=\columnwidth]{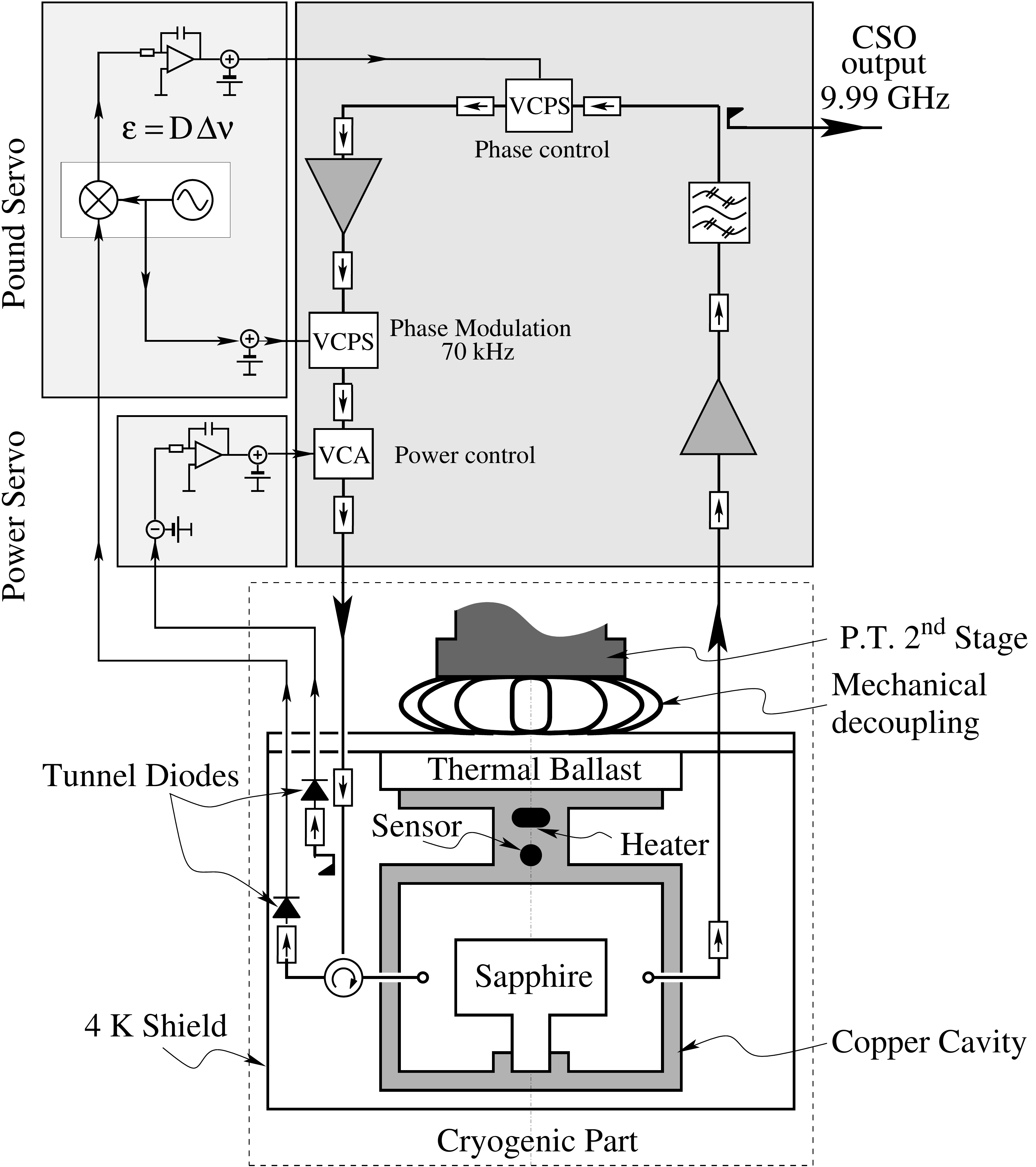}
	\caption{Scheme of the Cryogenic Sapphire Oscillator showing the cryogenic part (cooled resonator) and the electronics placed at room temperature}.
	\label{fig:fig1}
	\end{figure}

Our most advanced CSOs incorporate a $54$ mm diameter and $30$ mm height cylindrical sapphire resonator. It operates on the quasi-transverse magnetic whispering gallery mode $WGH_{15,0,0}$ near 10 GHz  \cite{rsi10-elisa}. The Q factor can achieve $1\times10^{9}$  at the liquid-He temperature depending on the sapphire crystal quality and on the resonator adjustment and cleanning.
In an autonomous cryocooled CSO, the sapphire resonator is placed into a cryostat and in thermal contact with the second stage of a pulse-tube (PT) cryocooler delivering typically 0.5 W of cooling power at 4 K (see Fig. \ref{fig:fig1}).  
The gas flow in the cryocooler induces mechanical vibrations and a temperature modulation at about 1 Hz, which need to be filtered. In our cryostats the heat-links between the PT 2$^{\mathrm{nd}}$ stage and the flange supporting the $4$ K thermal shield and the resonator is made with copper braids or foils. The same mechanical decoupling is implemented between the PT 1$^{\mathrm{st}}$ stage and the 50 K thermal shield.  This simple arrangement is sufficient to limit the resonator displacement below $1$ $\mu$m at the PT cycle frequency \cite{ell10-elisa}. The thermal  filtering is obtained by combining the heat-link thermal resistance and the thermal mass of the $4$ K flange and its ballast that could be added. Eventually, the resonator is stabilized at its turnover temperature $T_{0}$, where its thermal sensitivity nulls at first order. $T_{0}$ depends on the residual paramagnetic impurities present in the sapphire crystal \cite{mtt-2015}, and thus is specific to each resonator. $T_{0}$ is typically between $5$ and $8$ K for a high-quality sapphire crystal.\\

The CSO is Pound-Galani oscillator \cite{galani84}. In short, the resonator is used in transmission mode in a regular oscillator loop, and in reflexion mode as the discriminator of the classical Pound servo \cite{pound46}. The sustaining stage and the control electronics are placed at room temperature. The insertion loss through the cryostat is  $\approx -30$ dB. The sustaining stage is made up of commercial components. Two low noise microwave amplifiers provide a small signal gain of c.a. $54$ dB. A Voltage Controlled Attenuator (VCA) allows the control of the power injected in the resonator. Two Voltage Controlled Phase Shifters (VCPS) are used for the Pound servo. A 70 kHz phase modulation is applied through the first one. The correction is applied through the second VCPS. A $80$ MHz bandwdith filter and some isolators complement the circuit. The error signals needed for the Pound and the power servos are derived from the low frequency voltages generated by two tunnel diodes placed near the resonator input port. The Pound detector is directly connected to a lockin amplifier (Model Stanford Research Systems SR 810).

Three oscillators were assembled successively since 2012. They are identical in the principle but show however some differences (see table \ref{tab:table-1}). 

The thermal ballast is a piece of stainless steel placed between the PT $2^{nd}$ stage and the resonator support. Associated with the thermal resistance of the link, it is equivalent to a first order filter with a time constant $\tau_{B}$. 

$D$ is the Pound servo frequency discriminator sensitivity in (V/Hz). Formally, $D$ is taken at the demodulator output (see Fig. \ref{fig:fig1}). For slow frequency fluctuations $\Delta \nu (t)$, the error signal at the demodulator output is $\epsilon (t)= D \Delta \nu (t)$.  $D$ is experimentaly determined by applying an offset at the lockin amplifier output and measuring the resulting frequency shift. \\

\begin{table}[h!!!!!]
\centering
\caption{Cryogenic Sapphire Oscillators main characteristics.}
\begin{tabular}{|l|ccc|}
\hline
							& CSO-1			& CSO-2 			&	CSO-3 \\
\hline
\bf{Resonator}\rm				& && \\
Frequency $\nu_{0}$ (GHz)		& 9.988 		& 9.995 			&	9.987 \\
Material						& HEMEX			& HEMEX			& Kyropoulos	\\
Loaded Q-factor $Q_{L}$						& $1\times 10^{9}$	& $350\times 10^{8}$	& $400\times 10^{8}$ \\
Input coupling coef. $\beta_{1}$&		1		&		1		& 0.92\\
Turnover temperature $T_{0}$						& 6.238 K			&  5.766 K		& 6.265 K\\
\bf{Cryostat}\rm				&&&\\
Cryocooler Model				& PT 405			& PT 405			& PT 407 \\
Ballast time constant $\tau_{B}$	& 12 s				& 100 s			&	35 s\\
\bf{Control electronics} &&&\\
Pound Discri. Gain $D$	&$3.4$ mV/Hz & $2.3 $ mV/Hz & $1.4 $ mV/Hz\\	
Injected power 			& 100 $\mu$W			&	300 $\mu$W			&  70 $\mu$W\\
\hline
\end{tabular}
\label{tab:table-1}
\end{table} 

CSO-1 is an optimized copy of our first demonstrator, i.e. ELISA developed for the European Space Agency (ESA) and implemented in the Deep Space Network station DSA-3 in Malargue Argentina \cite{rsi10-elisa}. The second one ULISS is a transportable unit, which has been travelled since 2012 in some european laboratories to be tested in real field applications \cite{RSI-2012}. The third CSO has a new designed cryostat and was put into operation in october 2014. No completly optimized, it incorporates a crystal manufactured by the Kyropoulos growth method instead of a HEMEX crystal.\\

\section{Relative frequency stability measurements}

\begin{figure}[ht!]
	\centering
	\includegraphics[width=0.9\columnwidth]{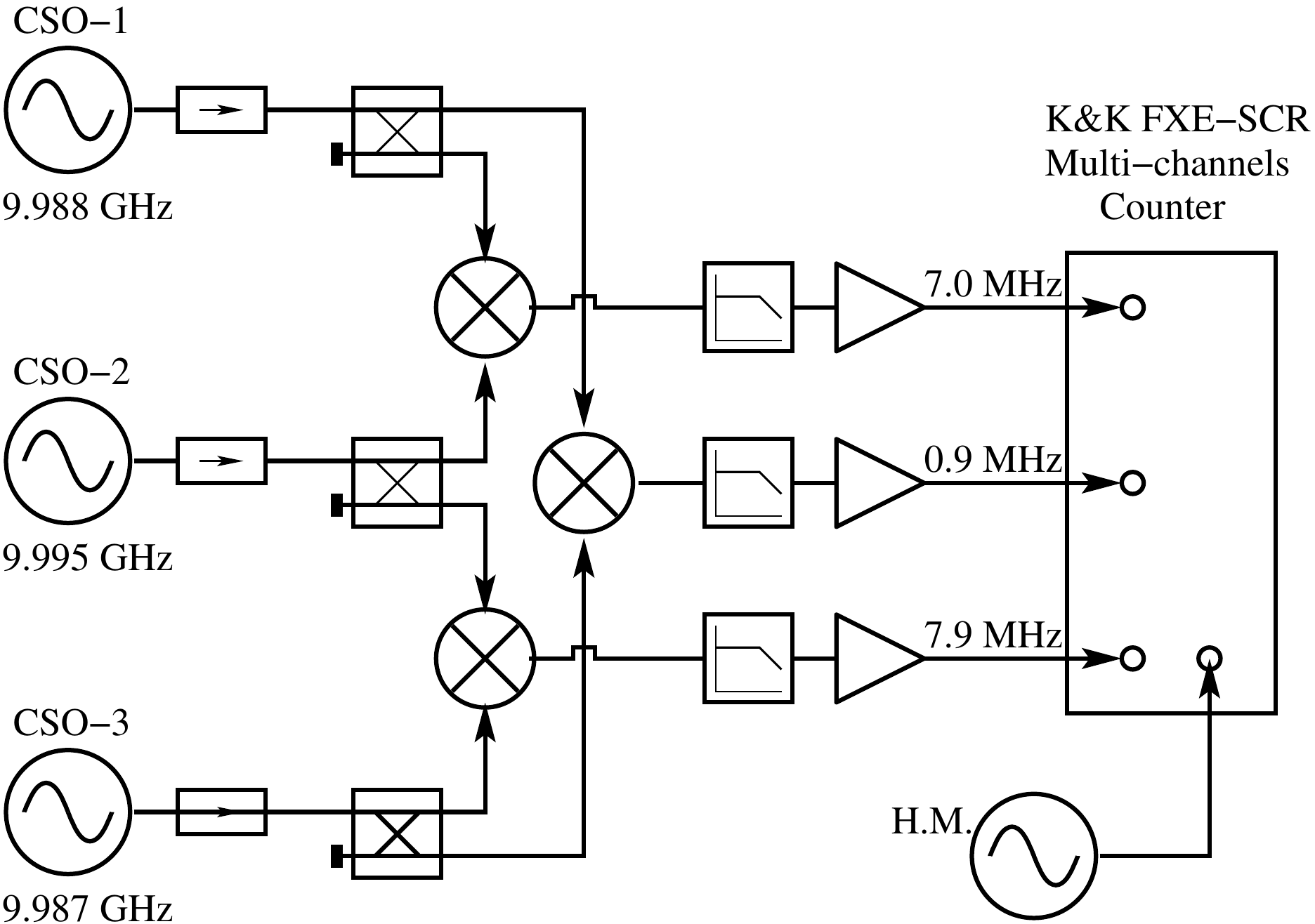}
	\caption{Measurement Set-Up. Each beatnote is low-pass filtered (10.9 MHz) and amplified (Minicircuits ZFL-1000+ Amplifier). The multichannel counter is referenced on the 10 MHz coming from an Hydrogen Maser (H.M.)}.
	\label{fig:fig2}
	\end{figure}

The measurement set-up is schematized in figure \ref{fig:fig2}. The CSO output signals are mixed to obtain the three beatnotes: $\nu_{12}=7$ MHz, $\nu_{13}=0.9$ MHz and $\nu_{23}=7.9$ MHz. They are simply counted using a multi-channels K$\&$K-FXE SCR counter \cite{kramer2001}. The three channels work in parallel thereby the data acquisitions are synchronous. All datas are post-processed by one second averaging time using a $\Pi$ windowing \cite{rubiola-rsi05}.

\subsection{Beatnotes and phase noise}
\begin{figure}[ht!]
	\centering
	\includegraphics[width=\columnwidth]{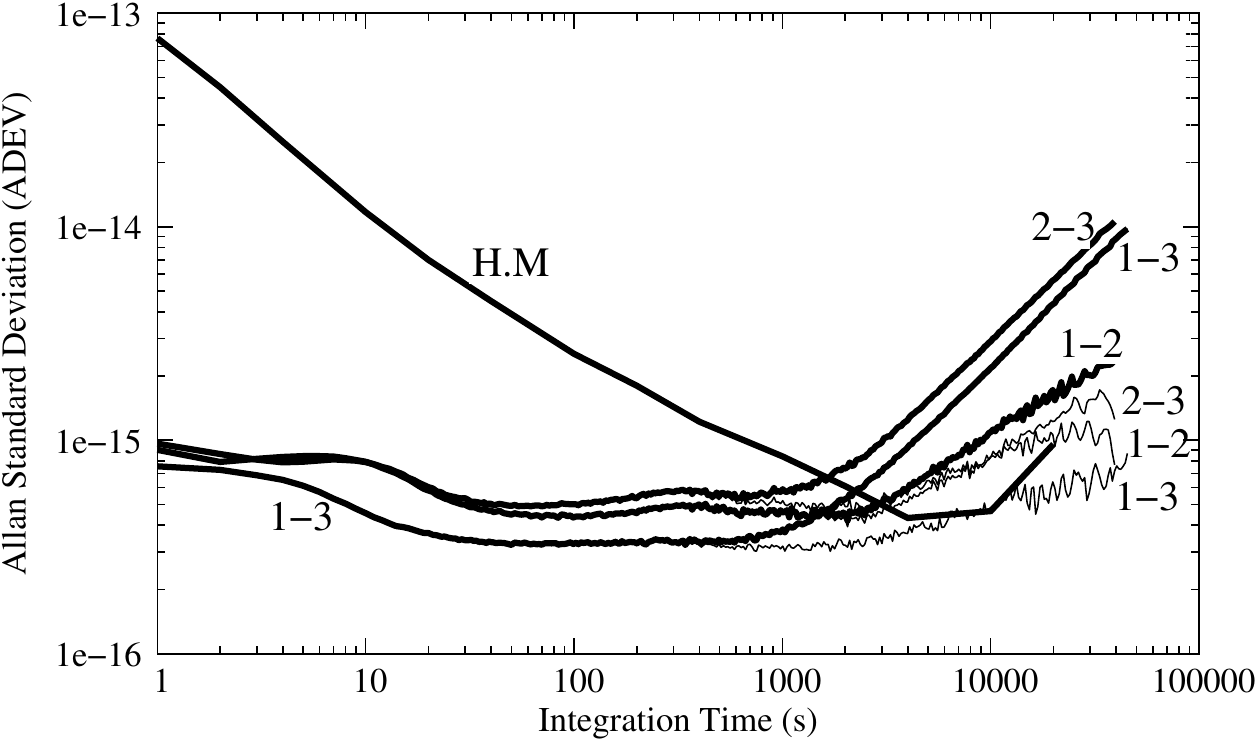}
	\caption{Allan Standard Deviations (ADEV) calculated from the three beatnotes.}
	\label{fig:fig3}
	\end{figure}

In figure \ref{fig:fig3}, the Allan standard deviations (ADEV) computed from the three beatnotes are comparated to the fractional stability of a typical high-performance hydrogen Maser (H.M.) The bold lines represente the ADEVs without any post data processing. The thin lines are the ADEV computed from the data after a linear drift removing. 

\begin{figure}[ht!]
	\centering
	\includegraphics[width=\columnwidth]{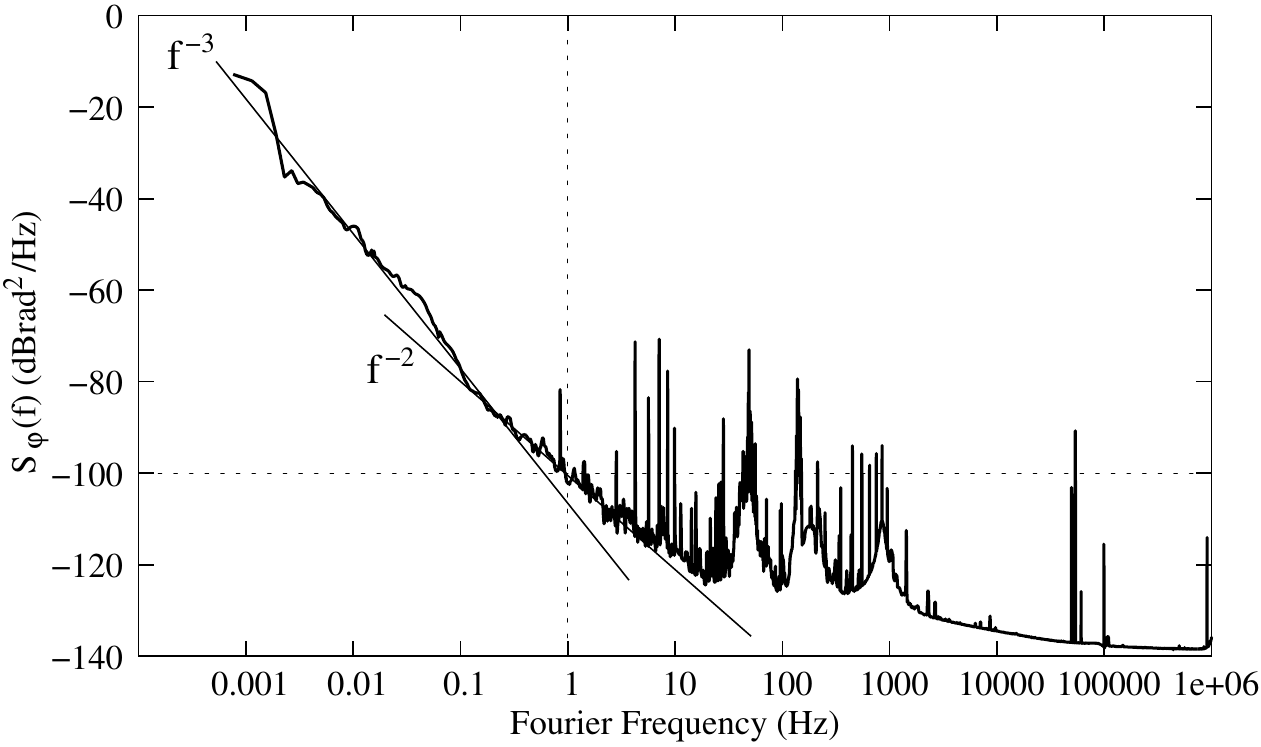}
	\caption{One unit phase noise obtained by comparing CSO-2 and CSO-3. Measurement realized with the Symmetricom 5125A phase test set-up.}
	\label{fig:fig4}
	\end{figure}

The measured ADEVs are better than $1\times 10^{-15}$ for $\tau \leq 2,000$ s. If the oscillator noises are equivalent and uncorrelated, that means that the fractional frequency stability of one unit is better than $7 \times 10^{-16}$, which is a conservative value as we will see in the next section. This number is coherent with the phase noise measurement between CSO-2 and CSO-3 shown in figure \ref{fig:fig4}.

This result corresponds to the phase noise of one unit assuming equivalent and uncorrelated the two CSOs: 3 dB has been substracted from the measured spectrum. For Fourier frequency $f\leq 10$ Hz, the phase noise spectrum can be approximated by a white frequency noise ($f^{-2}$ slope) and a flicker frequency noise ($f^{-3}$ slope): 

\begin{equation}
S_{\varphi}(f) \approx  ( \dfrac{1}{f^{-2}}+\dfrac{0.2}{f^{-3}}  ) \times 10^{-10} ~~\mathrm{rad}^{2} \mathrm{Hz}^{-1}
\end{equation}
This phase noise spectrum is equivalent in the time domain to an Allan standard deviation such as  \cite{rubiola-phase-noise}:
\begin{equation}
\sigma_{y}(\tau) = 7.1\times 10^{-16} \tau^{-1/2} + 5.3 \times 10^{-16}
\end{equation}

CSO-1 and CSO-3 present the best performances at short term, whereas a perturbation near $\tau=10$ s affects CSO-2.  The long term behaviors also differ: CSO-1 and CSO-2 do not show a frequency drift but seems limited by a randow walk process. Conversely CSO-3 is drifting with a rate of $2-3\times 10^{-14}$/day.

\subsection{Three Cornered Hat Method}	

\begin{figure}[ht!]
	\centering
	\includegraphics[width=\columnwidth]{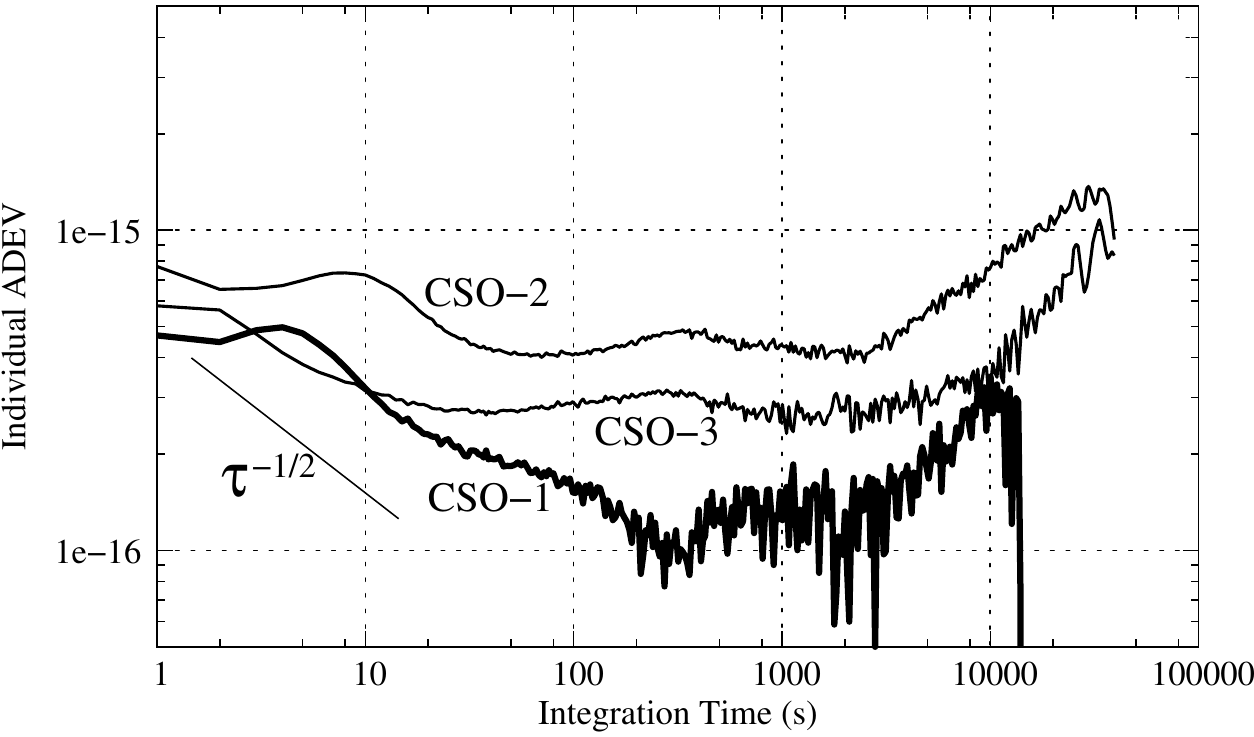}
	\caption{Individual ADEVs obtained by applying the three-cornered-hat method.}
	\label{fig:fig5}
	\end{figure}

The individual ADEVs have been computed using the three-cornered-hat method implemented in the Stable32 software. The results are given the figure \ref{fig:fig5}.

At short term ($\tau<\! 50 \mathrm{ s}$), all the individual ADEVs improve with the integration time but do not follow the expected $\tau^{-1/2}$ slope. Indeed it is expected that the CSO short-term frequency stability is ultimatly limited by the Pound discriminator white frequency noise. For CSO-1 and CSO-2 this white frequency noise is completely masked by another process leading to a hump in the ADEV. In section  \ref{sec:reso-temp-variations} we show this perturbation comes from the noise of the resonator temperature controller. At longer integration times, the ADEVs reach an apparent flicker floor expanding over approximatly two decades. Thus for the best oscillator, i.e. CSO-1, $\sigma_{y}(\tau) \approx 1.5 \times 10^{-16}$ for $100 \mathrm { s}\leq \tau \leq 5,000 \mathrm { s}$.\\

The humps appearing in the three ADEV at $\tau \sim400$ s, can be attributed to an oscillation in the air conditionning system of the lab. Thus they reveal the residual sensitivity to the room temperature of each CSO. Nevertheless the resulting CSO frequency variations are obviously correlated, which is in conflict with the requirement of the three-cornered-hat method. The calculated ADEV near 400 s cannot be considered as the actual frequency stability. Correlated noises indeed induces false results: the inversed hump in the CSO-1 ADEV is symptomatic of this situation. The same caution must be taken in the analysis of the long term fluctuations resulting also for a large part from the room temperature variations.

\section{Short term frequency stability Analysis}
\subsection{Noise in the resonator temperature control}
\label{sec:reso-temp-variations}
\begin{figure}[ht!]
	\centering
	\includegraphics[width=\columnwidth]{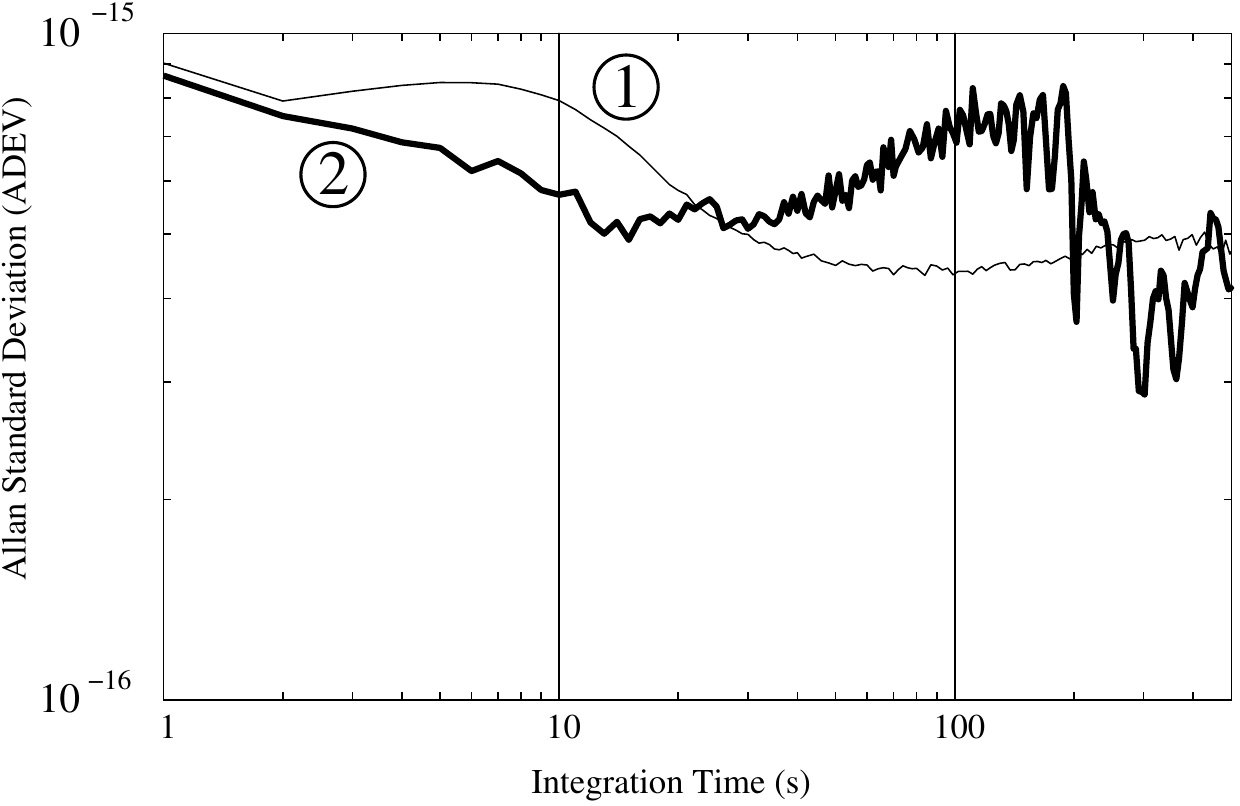}
	\caption{ADEV calculated from the beatnote CSO-1 vs CSO-2, for two temperature controller configuration: 1) reading in sensor unit ($\Omega$), 2) reading in tempearture unit (K).}.
	\label{fig:fig6}
	\end{figure}

The resonator is stabilized at its turnover temperature $T_{0}$ using a LakeShore Model 340 Temperature controller in the Proportional-Integral (PI) mode. The temperature sensor is a Cernox type CX-1050 with a sensitivity of approximately $3$ k$\Omega /$K and the actuator is a $25~\Omega$ heater. 
The LakeShore controller offers a variety of digital processing that can be done to the raw sensor data before applying the PI control equation. Thus the input information can be in sensor unit ($\Omega$) or converted in temperature unit (K). In this case a temperature response curve is needed to convert a sensor reading in sensor units to temperature. The configuration affects both the measurement resolution and the gain of the controller. 
The input data processing chosen, the setting of PI-controller gains is done using the autotuning procedure of the LakeShore controller. Eventually, the proportional and integral gains can be sighlty varied by checking the short term ADEV and searching for the best result. Figure \ref{fig:fig6} shows the ADEV calculated from the beatnote between CSO-1 and CSO-2 for two different configurations of the CSO-2 temperature controller:  reading in sensor unit ($\Omega$) or in temperature unit (K). The same proportional and integral gains are used for both. The effect on the frequency stability is obvious: the hump maximum is shifted to longer integration times when the reading is made in Kelvin. The postprocessing applied to the raw sensor data reduces the measurement resolution as well as the gain of the controller. 	

\begin{figure}[ht!]
	\centering
	\includegraphics[width=\columnwidth]{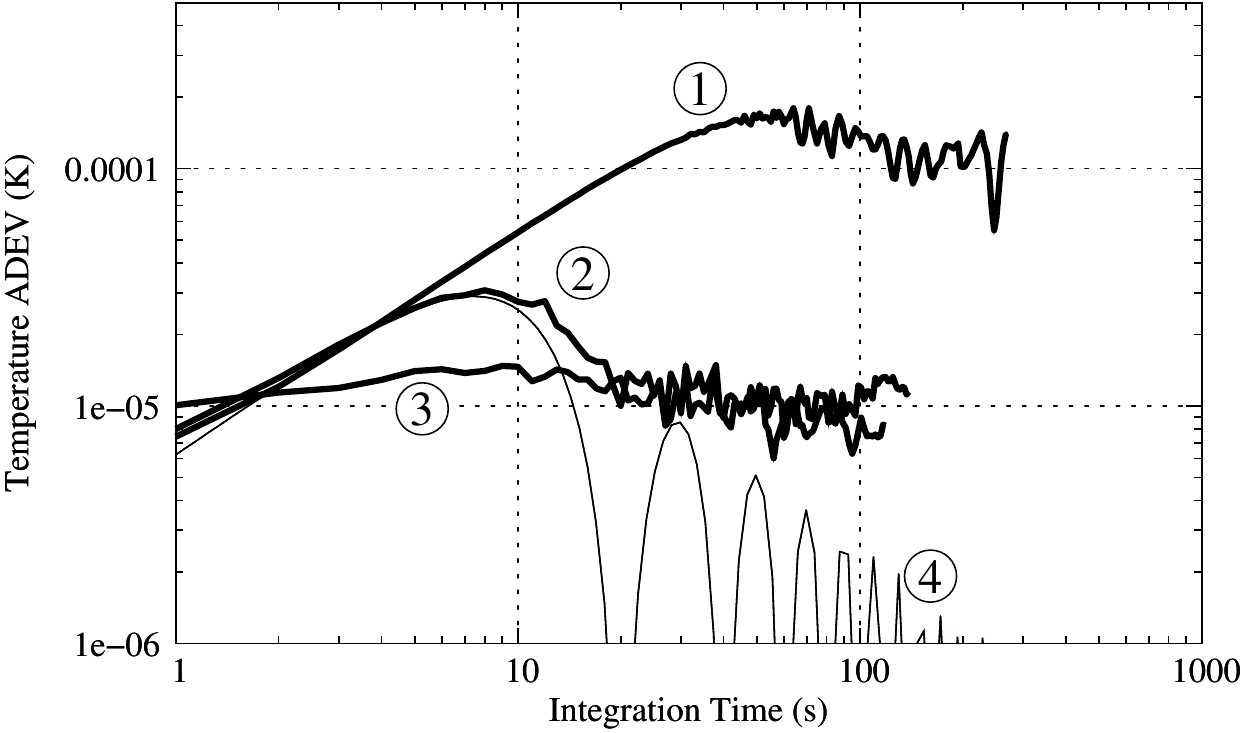}
	\caption{CSO-1 resonator temperature ADEV $\sigma_{T}(\tau)$ for some  temperature controller configurations, the proportional gain is kept constant: 1) reading in sensor unit ($\Omega$), integrator time constant $\tau_{I}=16$ s,  2) reading in tempearture unit (K), $\tau_{I}=16$ s, 3) reading in temperature unit (K), $\tau_{I}=8$ s, 4) Computed ADEV for a time varying temperature as: $T(t)=\langle T\rangle+\Delta T sin(2\pi\frac{t}{\theta_{m}})$ with $\Delta T=4\times 10^{-5}$ K and $\theta_{m}=20$ s.}
	\label{fig:fig7}
	\end{figure}

In a second time, we intentionnaly operate CSO-1 far from its turnover temperature, i.e at $T=T_{0}+100$ mK. At this point the first order resonator temperature sensitivity is  $\frac{1}{\Delta T}\frac{\Delta \nu}{\nu_{0}}=1.9\times10^{-10}/$K. Thus the CSO frequency fluctuations will follow those of the resonator temperature. Figure \ref{fig:fig7} shows the resonator temperature ADEV, i.e.  $\sigma_{T}(\tau)$, for different temperature controller configurations.

The curve $\textcircled4$ is the computed temperature ADEV assuming a resonator temperature modulated around its mean value $\langle T\rangle$ with a period $\theta_{m}=20$ s and an amplitude $\Delta T= 4\times 10^{-5}$ K, which represents well the actual behavior of the configuration $\textcircled2$. The curve $\textcircled3$ corresponds to the best configuration we found and that has been used for the measurements presented in the previous section. The curve $\textcircled3$ is just above the resolution measurement limited by the frequency noise of CSO-2. The temperature modulation is still there leading to the hump in the CSO-1 ADEV (see figure \ref{fig:fig5}). In spite of all our efforts we did not manage to find a better tuning for CSO-1 and CSO-2. CSO-3 behaves better as no hump has been observed in its ADEV.\\

The cause of this temperature modulation as well as its impact on the CSO frequency stability, are still not well understood. A modulation amplitude as $\Delta T=4\times10^{-5}$ K will lead to a frequency instability of $5\times 10^{-16}$ if the resonator temperature is $6$ mK above or bellow its turnover value $T_{0}$. Such an error on the temperature setpoint seems not realistic as it is much higher than the controller measurement resolution of $100~\mu$K. Moreover we tried to adjust the temperature setpoint by step of $1$ mK around the expected $T_{0}$ without finding any better tuning. 
The temperature modulation could result from an unexpected time lag in the thermal system making the control loop unstable. CSO-1 and CSO-2 are operating for a long time. They  have been subjected to several stops, and both have been transported by car in the frame of the ULISS project. Thermal anchorages and bold tightenings into the cryostat could have been degraded by the resulting mechanical perturbations. That is coherent with the fact that CSO-3 is immune to these thermal perturbation as it was recently assembled and has not been transported.

\subsection{Noise in the Frequency discriminator}
\label{sec:discri-noise}

\begin{figure}[ht!]
	\centering
	\includegraphics[width=\columnwidth]{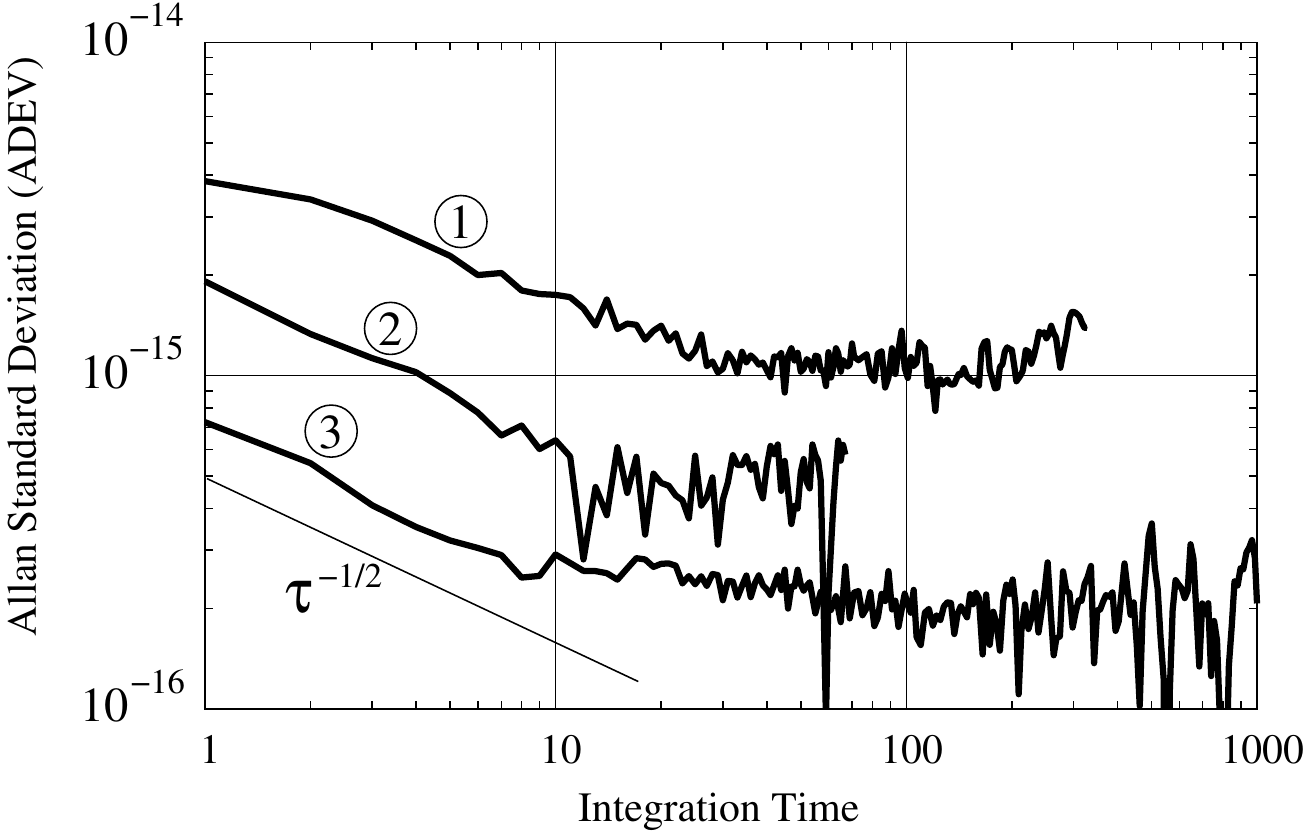}
	\caption{CSO-3 ADEV extracted from three-cornered-hat method. The frequency discriminator gain $D$ being 1) $1.3\times10^{-4}$ V/Hz, 2) $3.5\times10^{-4}$ V/Hz, 3) $1\times10^{-3}$ V/Hz.}
	\label{fig:fig8}
	\end{figure}

As explained in the previous section, CSO-3 is the only one not limited at short term by the resonator temperature fluctuations. Its ADEV is shown in figure \ref{fig:fig8} for several values of the Pound frequency discriminator gain $D$. The latter has been simply varied by changing the power $P_{R}$ injected into the resonator, all the other parameters being kept constant. 	

The measured ADEVs follow a $\tau^{-1/2}$ slope until $\tau \approx 30$s and then a flicker floor. Both vary with the incident power indicating that the CSO-3 performance is limited by the intrinsic noise of the Pound frequency discriminator. In presence of a white voltage noise at the demodulator output, the CSO short term fractional frequency stability is:
\begin{equation}
\sigma_{y}(\tau)=\dfrac{e_{n}}{\sqrt{2} D~ \nu_{0}} \tau^{-1/2}
\label{equ:sigma}
\end{equation}
where $e_{n}$ (V$/\sqrt{\mathrm{Hz}}$) is the demodulator output voltage spectral density. The computed values of $e_{n}$ are given in the table \ref{tab:en}.

\begin{table}[h!!!!!]
\centering
\caption{Computed voltage white noise spectral density: $e_{n}$}
\begin{tabular}{ccccc}
\hline
Curve in		& $\sigma_{y}(1\mathrm{s}) $ 	& $P_{R}$	& $D$				& $e_{n}$\\
Fig. \ref{fig:fig8}	&	$\times 10^{-15}$		&($\mu$W)	& $\times 10^{-4}$(V/Hz)	& (nV/$\sqrt{\mathrm{Hz}}$)\\
\hline\\
$\textcircled{1}$	&	5					& 12,5		&	1.3				&9.2 \\
$\textcircled{2}$	&	2					&33			&	3.5				& 9.9 \\
$\textcircled{3}$		&	0.7					&87			&	9.3				&9.2 \\	
\hline
\end{tabular}
\label{tab:en}
\end{table} 

The constant value $e_{n}\! \approx \! 9$ nV/$\sqrt{\mathrm{Hz}}$ is the equivalent voltage noise of the Pound detector. The direct measurement of $e_{n}$ requires to duplicate the Pound detector and thus to place a second cryogenic diode receiving the signal reflected by the cavity. This has not been foreseen in our current CSO design. However this value is compatible with the expected noise contributions of the Lockin amplifier and of the diode detector itself. \\

Assuming an identical Pound detector noise for CSO-1 and CSO-2 leads to a short term frequency stability of $3\times 10^{-16} \tau^{-1/2}$ and  $4.4\times 10^{-16} \tau^{-1/2}$ respectively. For this two CSOs the Pound discriminator noise is not the dominant process.

\section{Summary}
We applied the three-cornered-hat method to measure the individual fractional short-term frequency stability of three Cryogenic Sapphire Oscillators. This method implemented with commercial instruments and softwares permits a comparison at the $10^{-16}$ level. The method also reveals that there still exist technical sources of fluctuation responsible for a degradation of $\sigma_{y}(\tau)$ with respect to the Pound servo intrinsic noise. These perturbations could be minimized by a careful optimisation of the thermal system and of the resonator temperature stabilization. Despite these perturbations, all the tested CSOs present a short term frequency stability better than $7 \times 10^{-16}$ at 1 s and than $5 \times 10^{-16}$ between $30$ s and $3,000$ s.  The best CSO shows a frequency stability of $4.6\times 10^{-16}$ at $1$ s and a flicker floor below $2 \times 10^{-16}$.

\section*{Acknowledgments}

The work has been realized in the frame of the ANR project: Equipex Oscillator-Imp. The authors would like to thank the Council of the R\'egion de Franche-Comt\'e for its support to the \it{Projets d'Investissements d'Avenir}\rm\ and the FEDER for funding one CSO.



\begin{thebibliography}{10}

\bibitem{santarelli99-prl}
G.~Santarelli, P.~Laurent, P.~Lemonde, A.~Clairon, A.~G. Mann, S.~Chang, A.~N.
  Luiten, and C.~Salomon, ``Quantum projection noise in an atomic fountain: A
  high stability cesium frequency standard,'' {\it Physical Review Letters},
  vol.~82, no.~23, pp.~4619-4622, June~1999.

\bibitem{dick95}
G.~J. Dick, D.~G. Santiago, and R.~T. Wang, ``Temperature compensated sapphire
  resonator for ultra--stable oscillator capability at temperatures above
  77{K},'' {\it IEEE Transactions on Ultrasonics, Ferroelectrics and Frequency
  Control}, vol.~42, no.~5, pp.~812--819, 1995.

\bibitem{rsi10-elisa}
S.~Grop, P.~Y. Bourgeois, N.~Bazin, Y.~Kersal\'e, E.~Rubiola, C.~Langham,
  M.~Oxborrow, D.~Clapton, S.~Walker, J.~De~Vicente, and V.~Giordano, ``{ELISA:
  A} cryocooled 10 {GHz} oscillator with $10^{-15}$ frequency stability,'' {\it
  Review of Scientific Instruments}, vol.~81, no.~2, pp.~025102(1-7), 2010.

\bibitem{ell11-elisa}
S.~Grop, P.-Y. Bourgeois, E.~Rubiola, W.~Sch\"afer, J.~{De Vicente},
  Y.~Kersal\'e, and V.~Giordano, ``Frequency synthesis chain for the {ESA} deep
  space network,'' {\it Electronics Letters}, vol.~47,no.~6, pp.~386--388, Mar.~17,
  2011.

\bibitem{nand2011-mtt}
N.~Nand, J.~Hartnett, E.~Ivanov, and G.~Santarelli, ``Ultra-stable very-low
  phase-noise signal source for very long baseline interferometry using a
  cryocooled sapphire oscillator,'' {\it IEEE Transactions on Microwave Theory and Techniques}, vol.~59, no.~11, pp.~2978--2986, Nov. 2011.

\bibitem{doeleman2011}
S.~Doeleman, T.~Mai, A.~E.~E. Rogers, J.~G. Hartnett, M.~E. Tobar, and N.~Nand,
  ``Adapting a cryogenic sapphire oscillator for very long baseline
  interferometry,'' {\it Publications of the Astronomical Society of the
  Pacific}, vol.~123, no.~903, pp.~pp. 582--595, 2011.

\bibitem{rioja2012}
M.~Rioja, R.~Dodson, Y.~Asaki, J.~Hartnett, and S.~Tingay, ``The impact of
  frequency standards on coherence in {VLBI} at the highest frequencies,'' {\it
  The Astronomical Journal}, vol.~144, no.~4, pp.~121 (1-11), 2012.

\bibitem{AppPhysB-2014}
V.~Dolgovskiy, S.~Schilt, N.~Bucalovic, G.~Di~Domenico, S.~Grop, B.~Dubois,
  V.~Giordano, and T.~S\"udmeyer, ``Ultra-stable microwave generation with a
  diode-pumped solid-state laser in the 1.5-$\mu$m range,'' {\it Applied
  Physics B}, vol.~116, no.~3, pp.~593--601, 2014.

\bibitem{takamizawa2014}
A.~Takamizawa, S.~Yanagimachi, T.~Tanabe, K.~Hagimoto, I.~Hirano, K.-I. Watabe,
  T.~Ikegami, and J.~Hartnett, ``Atomic fountain clock with very high frequency
  stability employing a pulse-tube-cryocooled sapphire oscillator,'' {\it IEEE Transactions on Ultrasonics, Ferroelectrics, and Frequency Control},
  vol.~61, no.~9,pp.~1463--1469, Sept. 2014.

\bibitem{RSI-2012}
V.~Giordano, S.~Grop, B.~Dubois, P.-Y. Bourgeois, Y.~Kersal\'e, E.~Rubiola,
  G.~Haye, V.~Dolgovskiy, N.~Bucalovicy, G.~D. Domenico, S.~Schilt, J.~Chauvin,
  and D.~Valat, ``New generation of cryogenic sapphire microwave oscillator for
  space, metrology and scientific applications,'' {\it Review of Scientific
  Instruments}, vol.~83,  no.~8, pp.~085113(1-6), 2012.

\bibitem{hartnett-2012-apl}
J.~G. Hartnett, N.~R. Nand, and C.~Lu, ``Ultra-low-phase-noise cryocooled
  microwave dielectric-sapphire-resonator oscillators,'' {\it Applied Physics
  Letters}, vol.~100, no.~18, pp.~183501(1-4), 2012.

\bibitem{gray74}
J.~Gray and D.~Allan, ``A method for estimating the freqeuncy stability of an
  individual oscillator,'' in {\it Proc 28th Ann. Symp. on Frequency Control.},
  Fort Monmouth, (NJ) USA, pp.~243--246, May~29-31, 1974.

\bibitem{ifcs-2015-cso}
C.~Fluhr, S.~Grop, T.~Accadia, A.~Bakir, B.~Dubois, Y.~Kersal\'e, E.~Rubiola,
  and V.~Giordano, ``Characterization of a set of cryocooled sapphire
  oscillators at the $10^{-16}$ level with the three-cornered hat method,'' in
  {\it Proc. of the 2015 joint conference IEEE International Frequency
  Control Symposium (IFCS) and European Frequency and time Forum (EFTF)},
  Denver (CO) USA, 12-16 April 2015.

\bibitem{ell10-elisa}
S.~Grop, P.~Y. Bourgeois, R.~Boudot, Y.~Kersal\'e, E.~Rubiola, and V.~Giordano,
  ``10 {GHz} cryocooled sapphire oscillator with extremely low phase noise,''
  {\it Electronics Letters}, vol.~46, no.~6, pp.~420--422, 18th~March 2010.

\bibitem{mtt-2015}
V.~Giordano, F.~Christophe, S.~Grop, and B.~Dubois, ``Tests of sapphire
  crystals manufactured with different growth processes for ultra-stable
  microwave oscillators,'' {\it IEEE Transactions on Microwave Theory and
  Techniques}, accepted for publication Nov. 2015.

\bibitem{galani84}
Z.~Galani, M.~Bianchini, R.~Waterman, R.~Dibiase, R.~Laton, and J.~Cole,
  ``Analysis and design of a single-resonator {GaAs} {FET} oscillator with
  noise degeneration,'' {\it IEEE Transactions on Microwave Theory and
  Techniques}, vol.~32, no.~12, pp.~1556--1565, Dec. 1984.

\bibitem{pound46}
R.~Pound, ``Electronic frequency stabilization of microwave oscillators,'' {\it
  Review of Scientific Instruments}, vol.~17, pp.~490--505, Nov. 1946.

\bibitem{kramer2001}
G.~Kramer and W.~Klische, ``Multi-channel synchronous digital phase recorder,''
  in {\it Proc. of  the 2001 IEEE International Frequency Control Symposium and PDA Exhibition, 2001. }, pp.~144--151, 2001.

\bibitem{rubiola-rsi05}
E.~Rubiola, ``On the measurement of frequency and of its sample variance with
  high-resolution counters,'' {\it Review of Scientific Instruments}, vol.~76, no.~5, pp. 054703 (1-6)
  2005.

\bibitem{rubiola-phase-noise}
E.~Rubiola, {\it Phase noise and frequency stability in oscillators}.
\newblock Cambridge University Press, 2008.
\newblock ISBN 978-0-521-88677-2.

\end{thebibliography}
\end{document}